\documentclass[prl,twocolumn,superscriptaddress]{revtex4}
\usepackage{graphicx}
\usepackage{epsfig}
\usepackage{amssymb,amsmath,amsfonts,hyperref}
\usepackage{overpic} 
\usepackage{wasysym}
\usepackage{latexsym}
\usepackage{array,multirow}
\usepackage{verbatim}
\usepackage[normalem]{ulem}
\usepackage{color}

\newcommand{\be}{\begin{eqnarray}}
\newcommand{\ee}{\end{eqnarray}}

\begin{document}
\title{Glueball Spectra from a Matrix Model of Pure Yang-Mills Theory}
\author{Nirmalendu Acharyya$^1$, A.~P.~Balachandran$^{2,3}$, Mahul Pandey$^4$, Sambuddha Sanyal$^{5}$, Sachindeo~Vaidya$^{4,6}$ \\
${}^1${\small Optique Nonlin\'eaire Th\'eorique, Universit\'e libre de Bruxelles (U.L.B.), CP 231, 1050 Bruxelles, Belgium}\\
${}^2${\small Physics Department, Syracuse University, Syracuse, New York 13244-1130, U. S. A.}\\
${}^3$ {\small The  Institute of Mathematical Sciences, C.I.T Campus, Taramani, Chennai 600113, India} \\
${}^4${\small Centre for High Energy Physics,  Indian Institute of Science, Bengaluru, 560012, India}\\
${}^5$ {\small International Centre for Theoretical Sciences, Tata Institute of Fundamental Research, Bengaluru 560089, India}\\
${}^6${\small Perimeter Institute for Theoretical Physics, Waterloo, Ontario N2L 2Y5, Canada}}
\begin{abstract}
We present variational estimates for the low-lying energies of a simple matrix model that approximates $SU(3)$ Yang-Mills theory 
on a three-sphere of radius $R$. By fixing the ground state energy, we obtain the (integrated) renormalization 
group (RG) equation for the Yang-Mills coupling $g$ as a function of $R$. This RG equation allows to estimate the masses of other glueball states, which we find to be in excellent agreement with lattice simulations. 
\end{abstract}

\maketitle
\paragraph{\textbf{Introduction:}}
Quantum chromodynamics or QCD, the theory that underlies strong interactions, is an interacting non-Abelian gauge theory. It is expected that the self-coupling of the gauge field leads to bound states called glueballs, which emerge as particle excitations and interact with hadrons. The glueball has thus always been a topic of theoretical interest, but has eluded experimental verification till date. Theoretical predictions of the physical properties of glueballs pose a substantial challenge since their origin is deep inside the nonperturbative regime of the gauge theory. Much work on the properties of these particles has therefore been numerical, and the current status of lattice results, summarized in \cite{Morningstar:1999ff}, 
reflects the culmination of several decades of intense computational effort. Subsequent inclusion of quarks to 
give meson and baryon masses has been very challenging, and progress in estimation of light hadron masses has been rather recent \cite{Durr:2008zz}.

A matrix model of $SU(N)$ Yang-Mills theory proposed recently in \cite{Balachandran:2014iya, Balachandran:2014voa} successfully captures an important non-perturbative aspect of the full quantum theory, namely, the non-trivial nature of the gauge bundle \cite{gribov,singer,narasimhan}. This model has been used to deduce a surprising connection between the impure nature of (colored) states and the non-trivial topology of the bundle, with possible 
implications for confinement.

\begin{figure}[h!]
\begin{center}
\begin{overpic}[width=0.95\hsize]{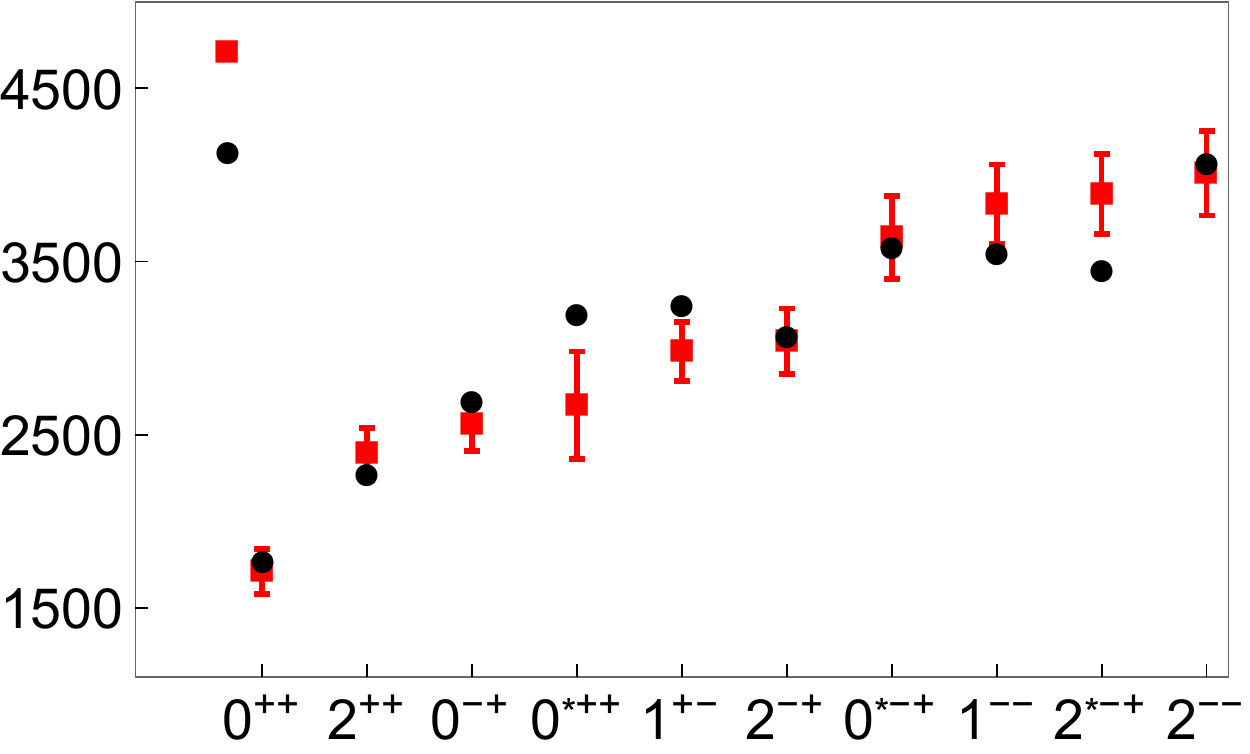}
\put(-5,15){\rotatebox{90}{Glueball Masses (MeV)}}
\put(40,-5){\rotatebox{0}{Glueball States}}
\put(20,56){\rotatebox{0}{Lattice QCD}}
\put(20,48){\rotatebox{0}{Matrix model}}
\end{overpic}
\end{center}
\caption{{The variational estimates of the glueball masses obtained from the matrix model of $SU(3)$ Yang Mills theory compared with the lattice QCD results (Morningstar and Peardon \cite{Morningstar:1999ff}, 1999)\label{fig_mass}.}}
\end{figure}
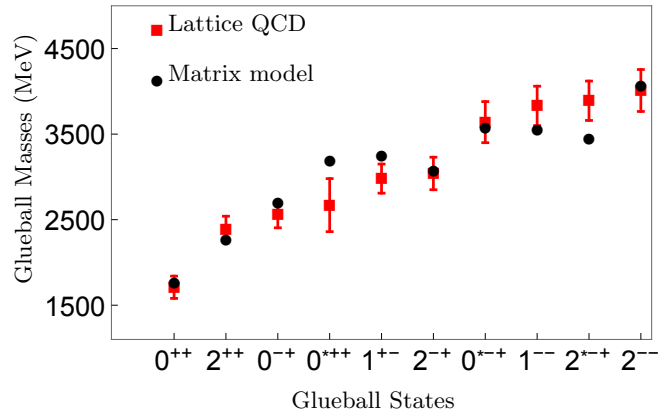

The model is based on a rectangular $3 \times (N^2-1)$ real matrix and avoids many technical difficulties of quantum field theory, and is amenable to numerical investigations.  In this letter, we demonstrate the extent to which realistic information about QCD may be extracted from this matrix model. Specifically, we compute the glueball masses from the low energy spectrum of the $SU(3)$ matrix model, and compare our results with the those from lattice simulations. Remarkably, we find that our results  match the lattice results quite accurately (within the lattice error bars), as summarized in Fig. \ref{fig_mass}. 

This is striking, as it indicates that the matrix model and the computation scheme which we present in this 
letter could be useful for QCD computations. Needless to say, the matrix model 
can never be an alternative to the field theory approach in its entirety.  However, a numerical analysis of this model 
provides a new tool for quick and easy estimation of the spectrum. 


Our matrix model Hamiltonian naturally takes the form $H = H_0 + V$ where $H_0$ is a $24$-dimensional harmonic oscillator, and the potential $V$ has cubic and quartic interaction terms. Using the eigenstates of $H_0$, we construct colorless wave functions  and apply standard variational techniques to estimate 
the low-energy spectrum of the Hamiltonian. 
A  renormalization prescription, detailed below, then allows us to relate this spectrum to glueball masses. 

In the rest of the letter we first briefly review the matrix model \cite{Balachandran:2014voa} and then give details of the calculation which lead to Fig. \ref{fig_mass}.
\paragraph{\textbf{The Matrix Model:}}
{A matrix model of the Yang-Mills theory can be constructed by compactifying the spatial $\mathbb{R}^3$ to 
$S^3$, and pulling back the Maurer-Cartan form on $SU(N)$ to this $S^3$ to obtain a particular subspace of the space of all gauge fields. Gauge fields in this subspace are $3 \times (N^2- 1)$ real matrices, yielding a $(0 + 1)$-dimensional matrix model of $SU(N)$ Yang-Mills theory. Below is a brief description of this procedure.}

The general left-invariant one-form on $SU(N)$ is $\Omega = \textrm{Tr}\left( T_a g^{-1} dg\right) M_{ab} T_b$ where $g \in SU(N)$  and $T_a$ are the Hermitian generators of the Lie algebra of $SU(N)$ in the fundamental representation. 
$M$ is a $(N^2-1)\times (N^2-1)$ real matrix, and the trace is in the fundamental representation of $SU(N)$. 

{The spatial-slice $S^3$ can be isomorphically mapped to an $SU(2)$ embedded in $SU(N)$. The action of left-invariant vector-fields $X_i$ of this $SU(2) \subset SU(N)$ yields $\Omega(X_i)=  -M_{ia} T_a.$
The spatial vector fields  are identified with $i X_i$. Thus the gauge fields on the spatial-slice are $A_0 =0, \quad A_i = -i M_{ia} T_a.$ The matrices $M_{ia}$ are the gauge variables of the matrix model. These $M$'s parametrize a submanifold of the space of all connections $\mathcal{A}$. 

Under a color transformation, $M \rightarrow M (Ad \,h)^T$, $h\in SU(N)$. The space $\mathcal{C}_N$ of inequivalent gauge configurations is $\mathcal{M}_N/Ad\, SU(N)$, where $\mathcal{M}_N$ is the space of all $3\times (N^2-1)$ real matrices.
In \cite{narasimhan}, it was shown $\mathcal{C}_2$ is a twisted bundle, and therefore no nonzero global 
section exists. 


%
%
The curvature $F_{ij}$ is obtained by pulling back the Maurer-Cartan equation to $S^3$: $F_{ij}^a= (d\Omega +\Omega \wedge \Omega) (i X_i, i X_j)= i(R^{-1} \epsilon_{ijk} M_{ka}- f_{abc} M_{ib} M_{jc})$
where $R$ is the radius of the $S^3$.

We re-scale  $M_{ia} \rightarrow g R M_{ia}$ so the $M$'s are dimensionless, and the coupling $g$ appears explicitly. The chromo-electric  field is $E_{ia}=\frac{d M_{ia}}{dt}$ and the chromo-magnetic field  is $B_{ia}=\frac{1}{2}\epsilon_{ijk}F_{jk}^a$. 
The Hamiltonian $H$ is given by
\begin{equation}
H= \frac{1}{2}(E_{ia}E_{ia} +B_{ia}B_{ia})=H_0 + \frac{1}{R} V_{int}(M).
\label{mmodel}
\end{equation}
Here 
\begin{equation}
H_0= \frac{1}{R} \left(- \frac{1}{2} \frac{\partial^2\,\,\,\,\,\,}{\partial M_{ia}^2} + \frac{1}{2} M_{ia} M_{ia} \right)
\end{equation}
is the Hamiltonian of a $3(N^2-1)$-dimensional harmonic oscillator, and
\begin{equation}
\begin{array}{ll}
V_{int} (M) = & -\frac{g}{2} \epsilon_{ijk} f_{abc}M_{ia} M_{jb}M_{kc} \\
& + \frac{g^2}{4} f_{abc}f_{ade} M_{ib} M_{jc}M_{id} M_{je}
\end{array}
\end{equation}
where $i=1,2,3$ and $a=1,\ldots N^2-1$.

In quantum theory, the Hamiltonian $H \equiv H(g)$ acts on the Hilbert space of functions $\psi(M)$ (with inner product 
 $\langle \psi_1 | \psi_2 \rangle = \int\prod_{i=1}^3 \prod_{a=1}^{N^2-1} dM_{ia} \bar{\psi}_1(M) \psi_2(M) $). 
 {The Gauss' law constraint in the matrix model implies that  
\begin{equation}
[G_a, \mathcal{O}] \equiv [f_{abc}E_{ib}M_{ic}, \mathcal{O}] =0, 
\end{equation}
where $\mathcal{O}$ is any observable. (So $\mathcal{O}$  is a first class variable in the sense of Dirac.)} 

The Hamiltonian (\ref{mmodel}) only takes into account the classical zero-mode sector of the full field theory. 
The full quantum field theory contributes an extra constant to this Hamiltonian, coming from the zero-point 
energy of all the higher, spatially dependent modes. To incorporate these zero-mode quantum effects correctly, 
we add to (\ref{mmodel}) a constant $C(R)$. The $R$ dependence comes from the fact that $C$ is the 
renormalized total zero-point energy (see for example \cite{DeGrand:1975cf}). 
To make the constant dimensionless, we write $C=\frac{c(R)}{R}$, and work with
\begin{equation}
H'=H+\frac{c(R)}{R}.
\label{fullH}
\end{equation}
We henceforth recognize (\ref{fullH}) as the true Hamiltonian $H$ and drop the prime.

\paragraph{\textbf{The Spectrum of $H$ and glueballs:}}
Angular momentum commutes with $H$, and hence the eigenstates and energies can be organized by the 
spin $s$. For a given $s$, we obtain a tower of energies $\mathcal{E}_n [s], n=0,1,\ldots$ of 
the form $\mathcal{E}_n [s]= \frac{f^{(s)}_n(g) +c(R)}{R}$, measured in units of $R^{-1}$. 

Energy differences depend on $g$ and $R$, but not on $c$. We emphasize that neither the bare coupling $g$ nor $R$ are directly measurable: in fact, masses of physical particles must be computed in the `flat space" limit $R \rightarrow \infty$. Taking such a limit at fixed $g$ makes all the $\mathcal{E}_n [s]$ vanish, very much like the naive scaling of lattice calculations. For 
meaningful non-trivial results, the bare coupling $g$ must depend on $R$ in such a manner that all physical quantities (in 
our cases the energies) have well-defined values at $R = \infty$. But energy must now be measured in some 
physical unit like MeV. If $x \equiv R/l$ is the radius of $S^3$ measured in these units, then 
{ $\mathcal{E}_n [s] = \left( \frac{f^{(s)}_n(g)}{x} +\frac{c(x)}{x} \right)\frac{1}{l}.$}
We can now make the bare coupling $g$ a function of $x$ such that a particular energy difference (say that of 
$2^{++}$ and $0^{++}$) is fixed to a constant, say, the observed value. Since glueballs have not yet been seen 
experimentally, we fix this using lattice results. {\it This function $g(x)$ is our integrated renormalization group equation}. 

In practice, it is easier to make $x$ a function of $g$:
\begin{equation}
x(g) = \frac{\mathcal{E}_0[2] - \mathcal{E}_0 [0]}{m(2^{++}) - m(0^{++})}
\label{rgeqn}
\end{equation}

The actual numerical values of the masses also need the asymptotic value of $c(x)/x$. To this end, we demand that 
the physical mass of our lowest glueball be fixed to the known lightest glueball mass. Again, since lattice 
computations are our only guide, we fix this to lie within the range predicted by lattice simulations. Using this $c(x)/x$, 
we can predict the masses of other glueball states.

We need to assign parity $P$ and charge conjugation $C$ to 
our variational eigenstates. Parity poses a minor challenge, since under $P$, the gauge field 
transforms 
as {$P: M_{ia}\rightarrow -M_{ia}$}
whereas the Hamiltonian $H(g)$ transforms to $H(-g)$ because of the cubic term. However, we find numerically that in the flat-space 
limit, the expectation value of the parity operator between the variational eigenstates asymptotes to $\pm 1$, 
and parity can then be assigned unambiguously. 

Under charge conjugation, { $C:M_{ia}T_a\rightarrow -M_{ia}T_a^*$}.
Charge
conjugation is a good symmetry of $H$ for all $g$, so the $C$ value can be assigned 
unambiguously for any $g$. 
\paragraph{\textbf{Computational Scheme:}}
To estimate the eigenvalues of (\ref{mmodel}) we consider trial wave functions that are colorless 
linear combinations of eigenstates $\Psi_{\{n_{ia}\}} $ of $H_0$. We denote them as $\psi_m^{s}$, where $s$ is the spin and $m$ a label for the trial functions.  The variational eigenvalues are simply the eigenvalues of the matrix $\widetilde{H}=\langle \psi_m^{s} | H | \psi_{m'}^{s'} \rangle $. 

Specifically, we consider $16$ states with spin-$0$, $10$ triplets with spin-$1$ and $18$ quintuplets with spin-$2$. All states in the same spin-1 (or spin-2) multiplet are degenerate, hence it suffices to consider only one state from each multiplet. Details of the states and a brief summary of the variational strategy are given in the supplementary material.





The only non-zero entries of $\widetilde{H}$ are $\langle \psi_{l}^{0} |H| \psi_{l'}^{0}\rangle$, $\langle \psi_{m}^{1} |H| \psi_{m'}^{1}\rangle$ and $\langle \psi_{n}^{2} |H| \psi_{n'}^{2}\rangle$ -- dubbed as
$\widetilde{H}^{0}$,$\widetilde{H}^{1}$ and $\widetilde{H}^{2}$ respectively. 
Hence
\begin{equation}
\widetilde{H}=\begin{pmatrix}
    \widetilde{H}^{0} &  &   \\
     & \widetilde{H}^{1} &   \\
     &  & \widetilde{H}^{2} 
  \end{pmatrix}.
\end{equation}

Expressing the cubic and quartic interaction terms in terms of the $24$-dimensional creation/annihilation 
operators gives each matrix element simply as a sum of products of delta functions. 
 
Once the matrix blocks of $\widetilde{H}$ are obtained, it is straightforward obtain the eigenvalues of $\widetilde{H}$ numerically.

\paragraph{\textbf{Results:}}
Ratios of various energy differences are independent of both $x(g)$ and $c(x)$. Numerically, we find that these ratios asymptote to constant values at a large $g$ (Fig. \ref{figure_rat_mass_diff}). This confirms the validity of 
our renormalization scheme: if we fix $x(g)$ using the lattice data for the difference of two glueball masses, and 
the asymptotic value of $c(x)/x$ for one of these masses, all other glueball masses asymptote to constant values at large $g$. {\it These asymptotic values are our predictions for the masses of the other glueballs.}

For $x(g)$ we use (\ref{rgeqn}) with $m(2^{++}) - m(0^{++})=500 \,\,\textrm{MeV}$. Fig. \ref{figR} shows $x(g)$ as a function 
of $g$.
\begin{figure}
\begin{overpic}[width=0.95\hsize]{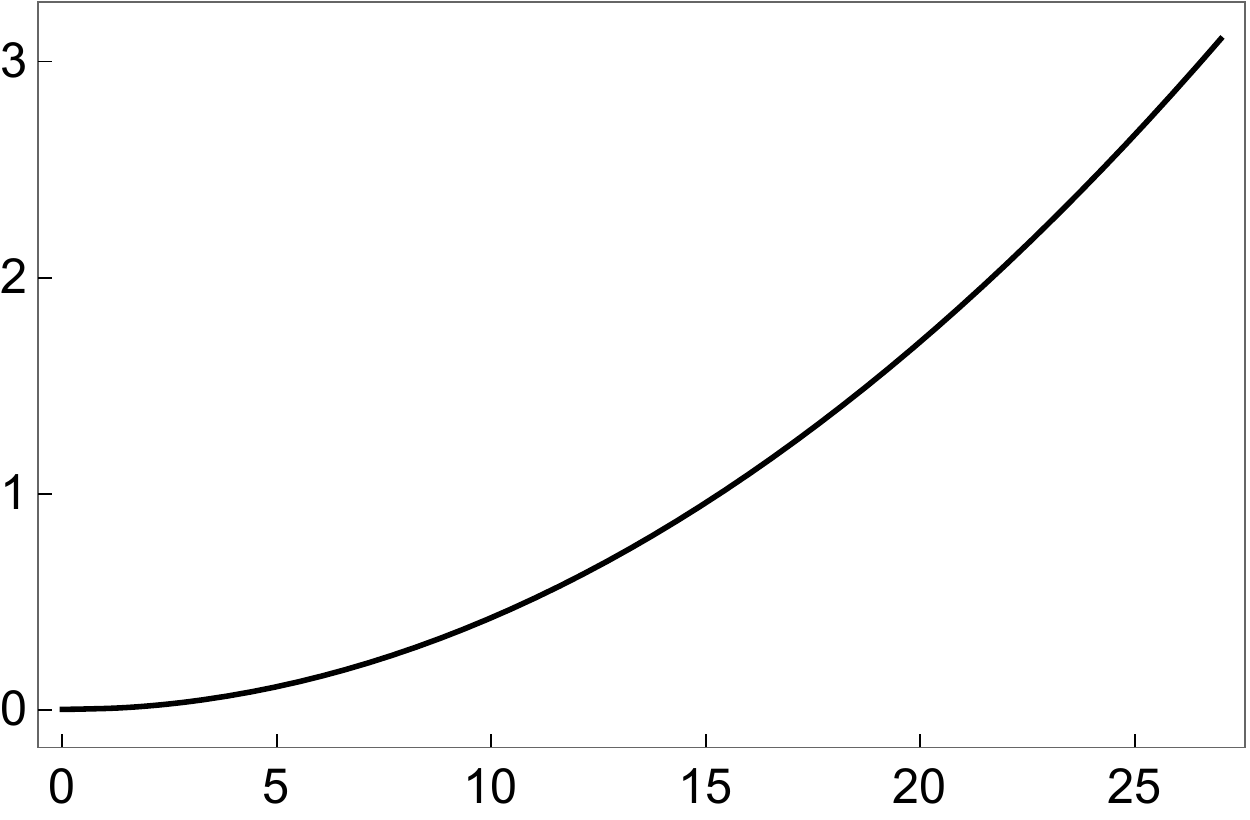}
\put(-6,30){\rotatebox{90}{$x(g)$}}
\put(50,-3){$g$}
\end{overpic}
\caption{RG flow equation for the Yang-Mills coupling $g$ as a function of $x$. It is shown here as $x(g)$ versus $g$ for convenience.}\label{figR}
\end{figure}

To set the asymptotic value of $\frac{c(x)}{x}$, we require that the physical mass of our lowest glueball lies within the range predicted by lattice simulations (1580 MeV--1840 MeV). Choosing this value to be {$1050$} MeV gives us the best fit with lattice predictions \cite{Morningstar:1999ff}. Our 
results for glueball masses, along with those from lattice simulations, are summarized in table \ref{Table_2}.
\begin{figure}
\begin{center}
\begin{overpic}[width=0.9\hsize]{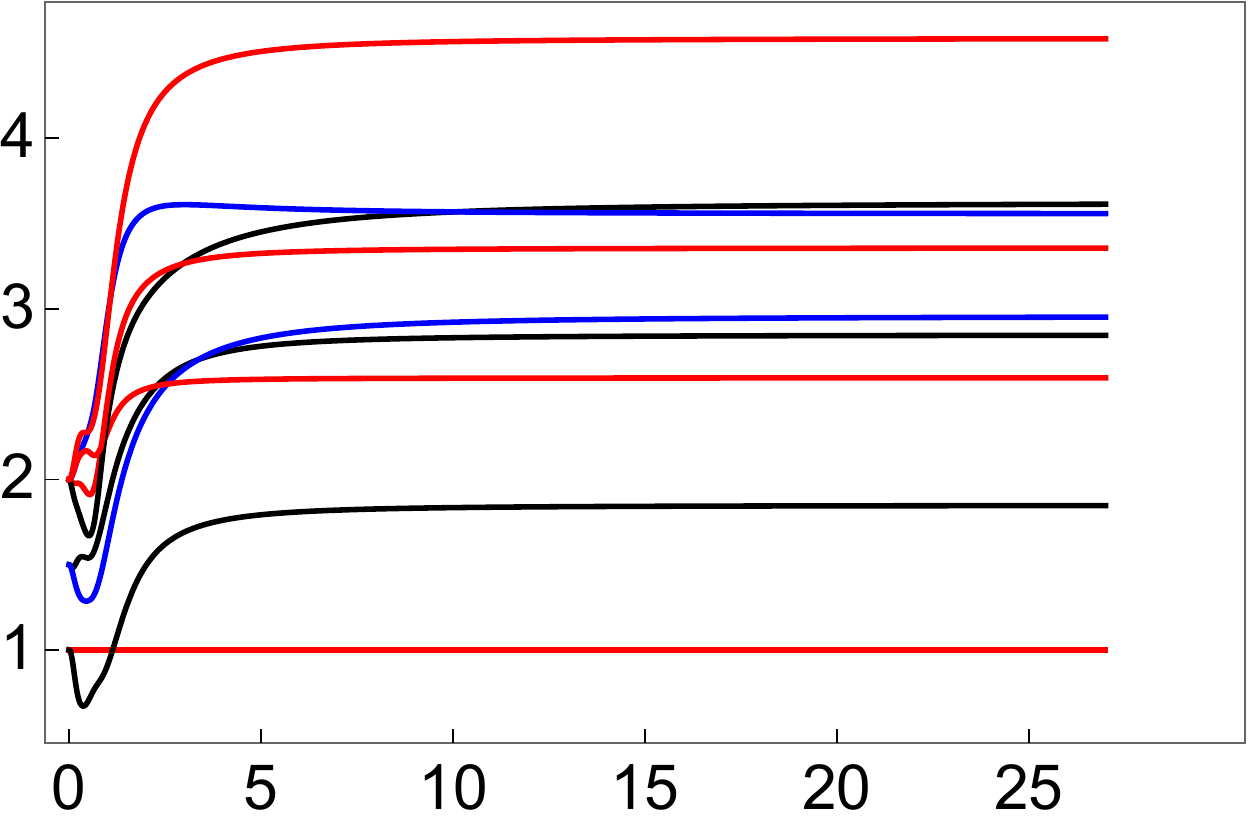}
\put(-9,20){\rotatebox{90}{$\frac{m(X) - m(0^{++})}{m(2^{++}) - m(0^{++})}$}}
\put(50,-3){$g$}
\put(90,10){$2^{++}$}
\put(90,22){$0^{-+}$}
\put(90,36){$0^{\ast++}$}
\put(90,39){$1^{+-}$}
\put(90,33){$2^{-+}$}
\put(90,44){$2^{\ast-+}$}
\put(90,47){$1^{--}$}
\put(90,50){$0^{\ast-+}$}
\put(90,62){$2^{--}$}
\end{overpic}
\caption{Ratios of mass differences $\frac{m(X) - m(0^{++})}{m(2^{++}) - m(0^{++})}$ as a function of $g$.  $X$ stands for the various glueballs shown in the figure. {(The black, blue and red curves represent spin-0, spin-1 and spin-2 levels respectively.)} }
\label{figure_rat_mass_diff}
\end{center}
\end{figure}

\begin{figure}[b!]
\quad\quad \begin{center}
 \begin{overpic}[width=0.95\hsize]{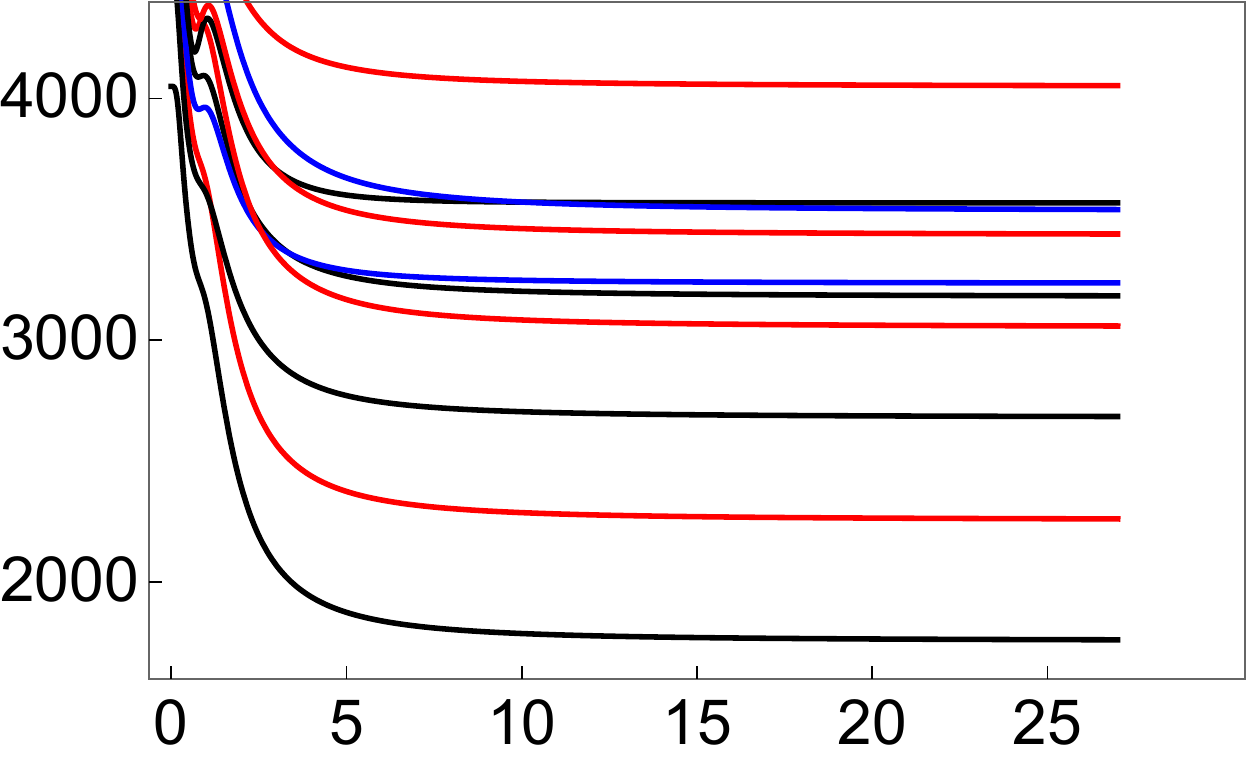}
\put(-5,10){\rotatebox{90}{Masses of glueballs (MeV) }}
\put(50,-4){$g$}
\put(90,8){$0^{++}$}
\put(90,19){$2^{++}$}
\put(90,27){$0^{-+}$}
\put(90,37){$0^{\ast++}$}
\put(90,40){$1^{+-}$}
\put(90,34){$2^{-+}$}
\put(90,43){$2^{\ast-+}$}
\put(90,46){$1^{--}$}
\put(85,49){{\small$0^{\ast-+}$}}
\put(90,57){$2^{--}$}
\end{overpic}
\caption{The glueball masses as a function of $g$.  Here we used $x(g)$ as in eqn. (\ref{rgeqn}).  {(The black, blue and red curves represent spin-0, spin-1 and spin-2 states respectively.)} }
\label{fig_mass_2}
\end{center}
\end{figure}
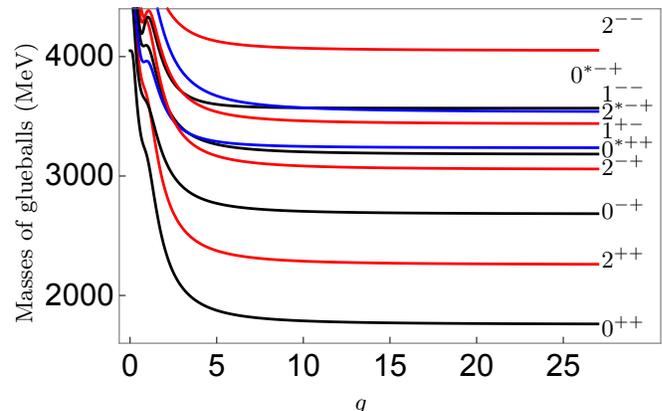
\begin{table}[h]
\begin{center}
\begin{tabular}{|c|c|c|}\hline&& \\
Glueball &Physical masses   & Physical masses\\ 
states & from matrix model &from lattice QCD \\$J^{PC}$&(MeV) & (MeV)\\ &&\\\hline&&\\
$0^{++}$ & {1757.08} & 1580 - 1840  \\ &&\\
$2^{++}$  &{2257.08} & 2240 - 2540 \\ &&\\
$0^{-+}$ & {2681.45} & 2405 - 2715  \\&&\\
$ 0^{*++}$&{3180.82} & 2360 - 2980   \\ &&\\
$ 1^{+-}$& {3235.41} & 2810 - 3150  \\  &&\\
$2^{-+}$ &{3054.97} & 2850 - 3230    \\&&\\
$ 0^{*-+}$ &{3568.02} & 3400 - 3880   \\ &&\\
$ 1^{--} $  & {3535.66}  &3600 - 4060 \\&&\\
$ 2^{*-+}$  &{3435.75} & 3660 - 4120  \\ &&\\
$ 2^{--}$ &{4050.14} & 3765 - 4255   \\  &&\\\hline

\end{tabular}\caption{{To obtain the estimates of the physical masses from the matrix model, $x(g)$ is determined by using the lattice data for the difference of two lowest glueball masses $m(2^{++}) - m(0^{++})=500 \,\,\textrm{MeV}$. The asymptotic value of $c(x)/x$ is fixed to $1050~ \textrm{MeV}$. Then all other glueball masses asymptote to constant values at large $g$ (Fig.~\ref{fig_mass_2}), which are our predictions for the masses of these
 glueballs. They are compared with their counterparts from Lattice QCD simulations \cite{Morningstar:1999ff}.} }\label{Table_2}
\end{center}
\end{table}

From Fig. \ref{fig_mass}, we can see that the matrix model predictions for the masses of $ 0^{*++}$ and $ 2^{*-+}$ 
fall outside the error bars of lattices simulations \cite{Morningstar:1999ff}. The lattice QCD mass estimates (and 
the error bars) of these very states are not very precise (see Fig. 8 in \cite{Morningstar:1999ff}). For $ 0^{*++}$, the lack of precision is due to the poor statistics near the continuum limit and consequently, the finite volume effects in the estimated mass are non-negligible. For $2^{\ast++} $, the difficulty in estimating the mass precisely is due to the presence of two other glueball states in the vicinity.
%
%

{ In Fig. \ref{fig_mass_2}, we observe that there are some level crossings at small $g$ between states of same 
spin. However, no physically significant information can be inferred from these crossings: the matrix model has a limited validity at small $g$ as the quantum field theoretic effects are expected to dominate in this regime.}

It is striking that the low lying glueball masses computed from the matrix model agree so well with lattice QCD 
simulations. On inclusion of more variational states, the  estimates are expected to improve further but at the cost of increased computational complexity. Compared to lattice simulations, the matrix model employs simpler numerical techniques and requires significantly less computing time. Our work demonstrates that the model can be deployed to make reasonably accurate estimates in the pure gauge sector, and may serve as a powerful and efficient numerical tool in the future study of  QCD.

%
\paragraph{\textbf{Outlook:}}

There has been a recent interest in estimating the glueball masses for large $N$ gauge theories using 
holographic models (for example, Witten-Sakai-Sugimoto Model) \cite{Brunner,Chen}. It is straightforward 
to generalize our computational techniques to such models. 

Recent work \cite{mahul1} shows that the matrix model coupled to fundamental fermions has 
superselection sectors, and the full Hilbert space breaks up into quantum mechanically disjoint subspaces. 
These sectors may be studied in detail using variational techniques similar to those presented here. The 
numerical study of the spectrum of this model is currently under way \cite{fermion_apsv}. An agreement 
with the experimental values of light hadron masses would provide a further convincing demonstration of 
the computational efficiency of 
the matrix model, and an important aid to detailed lattice QCD simulations \cite{Durr:2008zz,Durr:2008rw,Davies:2003ik, Aoki:2008sm,Bernard:2001av,Aubin:2004wf,Ukita:2007cu,Antonio:2006px,Alexandrou:2008tn,Noaki:2007es}.

\mbox{} \\
\textbf{Acknowledgements}\\
It is our pleasure to thank Manish Jain,  Apoorva Patel and especially Prasad Hegde for valuable discussions and inputs, Manolo Asorey and Fernando Falceto for emphasizing the importance of zero-point energy,  
and Sumati Surya and Gregory Kozyreff for a critical reading of the manuscript. A.P.B.  thanks the Institute of Mathematical Sciences, Chennai  for support during the completion of this work. S.S.  gratefully  acknowledges  funding  from DST  (India)  
and  DAE-SRC  (India)  and  support  from IISc (Bengaluru) during completion of part of this work.


\newpage
\onecolumngrid
\appendix
\section{Supplemental Material for ``Glueball Spectra in Matrix Model of Pure Yang-Mills Theory''}
In this Supplemental Material, we describe our variational scheme, provide the list of trial wave functions, and a brief summary of the numerics.

 \subsection{Variational calculation} 
 We construct the following variational ansatz 
\begin{equation}
|\chi\rangle = \sum_{i,s} c_i^s |\psi^s_i\rangle, \quad c_i^{s} \in \mathbb{C}. 
\end{equation}
To estimate the glueball masses, we consider the functional 
\begin{equation}
\mathcal{K}= \langle \chi |H | \chi \rangle - \lambda (\langle\chi|\chi \rangle -1)
\end{equation}
where $\lambda$ is a Lagrange multiplier for the constraint that the state $|\chi\rangle$ is normalized to 1. 

We minimize this functional wrt the parameters $c_i$ and $\lambda$;
\begin{equation}
\frac{\partial \mathcal{K}}{\partial \lambda} =0 =\frac{\partial \mathcal{K}}{\partial c_i^{s \ast}}
\end{equation}

Minimizing wrt $c_i^{s\ast}$ leads to the generalized eigenvalue equation,
\begin{equation}
\sum_{j,s^\prime}\widetilde{H}_{ij}^{s s^\prime}  c_j^{s^\prime} =\lambda \sum_{j,s^\prime}  \widetilde{S}_{ij}^{s s^\prime} c^s_j,
\label{evaleqn}
\end{equation}
where
\begin{equation}
\widetilde{H}_{ij}^{s s^\prime} \equiv \langle \psi_i ^{s} |H|\psi_j^{s^\prime}\rangle
\end{equation}
and
\begin{equation}
\widetilde{S}_{ij}^{s s^\prime} \equiv \langle \psi_i ^{s} |\psi_j^{s^\prime}\rangle.
\end{equation}

Minimizing wrt $\lambda$ yields
\begin{equation}
\sum_{ij,ss^\prime} c_i^{s*}\widetilde{S}^{ss^\prime}_{ij}c^{s^\prime}_j=0
\end{equation}

Multiplying on the left of (\ref{evaleqn}) by $c_k^{t*}$ and summing over $k$ and $t$ we obtain
\begin{align}
\lambda &=\frac{\sum_{jk,st}c_k^{t*}\widetilde{H}_{kj}^{ts}c^s_j}{\sum_{jk,st}c_k^{t*}\widetilde{S}_{kj}^{ts}c^s_j}\\
&=\frac{\langle \chi|H|\chi\rangle}{\langle \chi|\chi\rangle}.
\end{align}
So the lowest eigenvalue gives the ground state and the rest give the excited states.

It is not difficult to see that the matrix elements $\widetilde{H}_{ij}^{s s^\prime} = 0$ if $s\neq s^\prime$.  So $\widetilde{H}$ is block diagonal matrix of the form 
\begin{equation}
\widetilde{H}=\begin{pmatrix}
    \widetilde{H}^{0} &  &   \\
     & \widetilde{H}^{1} &   \\
     &  & \widetilde{H}^{2} 
  \end{pmatrix}
\end{equation}
Similarly, the Gram matrix $S_{ij}$ also has this block diagonal form.

\subsection{The Eigenstates of the Harmonic Oscillator and the  Colorless States  } 

$H_0$ is the Hamiltonian of a $3(N^ 2-1)$-dimensional harmonic oscillator for $SU(N)$: 
\begin{equation}
H_0= - \frac{1}{2R} \frac{\partial^2\,\,\,\,\,\,}{\partial M_{ia}^2} + \frac{1}{2R} M_{ia} M_{ia}, \quad\quad i=1,2,3,  \quad\quad a=1,2 \ldots (N^2-1).
\end{equation}

 $\Psi_{\{n_{ia}\}}$'s are the orthonormal eigenfunctions of the $3(N^2-1)$-dimensional harmonic oscillator 
\begin{equation}
H_0 \Psi_{\{n_{ia}\}} =\frac{1}{R}\left( n+\frac{3(N^2-1)}{2}\right) \Psi_{\{n_{ia}\}} ,\quad\quad n\equiv \sum_{i,a}n_{\{n_{ia}\}}.
\end{equation}

In terms of the matrix model variables $\{M_{ia}\}$, 
\begin{equation}
\Psi_{\{n_{ia}\}} = \prod_{i, a} \frac{1}{\sqrt{\sqrt{\pi} 2^{n_{ia}} n_{ia}}}H_{n_{ia}}(M_{ia}) e^{-\frac{1}{2} M_{ia}M_{ia}}.
\end{equation}

The discussion can be simplified by expressing the trial wavefunctions in terms of oscillator creation and annihilation operators. We define

\begin{equation}
A_{ia}=\frac{1}{\sqrt{2}}\left(M_{ia}+\frac{\partial}{\partial M_{ia}}\right); \hspace{5 mm} A^\dagger_{ia}=\frac{1}{\sqrt{2}}\left(M_{ia}-\frac{\partial}{\partial M_{ia}}\right).
\end{equation}
This relation implies
\begin{equation}
[A_{ia},A^\dagger_{jb}]=\delta_{ia}\delta_{jb}
\end{equation}

The oscillator vacuum is defined as
\begin{equation}
\langle M| 0\rangle=\frac{1}{\pi^{6}} e^{- \frac{Tr(M^T M)}{2}}
\end{equation}
where $|M\rangle\equiv |\{M_{ia}\}\rangle$ are the position eigenstates in the space of $\{M_{ia}\}$. Then a harmonic oscillator eigenstate with $n$ oscillators can be constructed by successive action of $A^\dagger_{ia}$ on the vacuum.

An $n$-oscillator state will in general transform under some representation of color. Physical states are color singlets, so we only consider colorless linear combinations of such states in our variational ansatz. They can be constructed by taking composites of the operators $A^\dagger_i\equiv A^\dagger_{ia}T_a$, where $T_a=\frac{\lambda_a}{2}$ are the generators of $SU(3)$, and tracing over the color. In our analysis, we have chosen all possible colorless states for $n\leq 6$. We denote them as $|\psi_i^s\rangle$, where the superscript $s$ is the spin of the state with respect to spatial rotations. We have considered spin-$0$, spin-$1$ and spin-$2$ states.

For spin $0$, we have a total of $16$ states, listed below:

\begin{eqnarray}
\begin{array}{lll}
|\psi_1^0 \rangle= |0\rangle\\
|\psi_2^0\rangle = A^\dagger_{ia}A^\dagger_{ia}|0\rangle\\
|\psi_3^0 \rangle= \epsilon_{ijk}f_{abc}A^\dagger_{ia}A^\dagger_{jb}A^\dagger_{kc}|0\rangle\\
|\psi_4^0\rangle = A^\dagger_{ia}A^\dagger_{ia} A^\dagger_{jb}A^\dagger_{jb}|0\rangle\\
|\psi_5^0 \rangle= A^\dagger_{ia}A^\dagger_{ib} A^\dagger_{ja}A^\dagger_{jb}|0\rangle\\
|\psi_6^0\rangle = d_{abe}d_{cde} A^\dagger_{ia}A^\dagger_{ib} A^\dagger_{jc}A^\dagger_{jd}|0\rangle\\
|\psi_7^0 \rangle= \epsilon_{ijk}f_{abc}A^\dagger_{ia}A^\dagger_{jb}A^\dagger_{kc} A^\dagger_{ld}A^\dagger_{ld}|0\rangle\\
|\psi_8^0\rangle = \epsilon_{ijk}f_{abc}d_{a_1b_1e}d_{a_2ce}A^\dagger_{ia}A^\dagger_{jb}A^\dagger_{ka_1} A^\dagger_{lb_1} A^\dagger_{l a_2}|0\rangle\\
|\psi_9^0 \rangle=  A^\dagger_{ia}A^\dagger_{ia} A^\dagger_{jb}A^\dagger_{jb}A^\dagger_{kc}A^\dagger_{kc}|0\rangle\\
|\psi_{10}^0\rangle =  A^\dagger_{ia}A^\dagger_{ib} A^\dagger_{jb}A^\dagger_{jc}A^\dagger_{kc}A^\dagger_{ka}|0\rangle\\
|\psi_{11}^0 \rangle= \epsilon_{ijk}\epsilon_{lmn} A^\dagger_{ia}A^\dagger_{la} A^\dagger_{jb}A^\dagger_{mb}A^\dagger_{kc}A^\dagger_{nc}|0\rangle\\
|\psi_{12}^0\rangle=\epsilon_{i_1j_1k_1}f_{a_1b_1c_1}\epsilon_{i_2j_2k_2}f_{a_2b_2c_2}A^\dagger_{i_1a_1}A^\dagger_{j_1b_1}A^\dagger_{k_1c_1}A^\dagger_{i_2a_2}A^\dagger_{j_2b_2}A^\dagger_{k_2c_2}|0\rangle\\
|\psi_{13}^0 \rangle= d_{abc}d_{def} A^\dagger_{ia}A^\dagger_{id} A^\dagger_{jb}A^\dagger_{je}A^\dagger_{kc}A^\dagger_{kf}|0\rangle\\
|\psi_{14}^0\rangle =d_{b_1c_1d}d_{b_2c_2d}A^\dagger_{ia}A^\dagger_{ia} A^\dagger_{jb_1}A^\dagger_{jc_1}A^\dagger_{kb_2}A^\dagger_{kc_2}|0\rangle\\
|\psi_{15}^0 \rangle= \epsilon_{i_1j_1k_1}f_{a_1b_1c_1}\epsilon_{i_2j_2k_2}f_{a_2b_2c_2}d_{c_1d_1e}d_{c_2d_2e}A^\dagger_{i_1a_1}A^\dagger_{j_1b_1}A^\dagger_{k_1d_1}A^\dagger_{i_2a_2}A^\dagger_{j_2b_2}A^\dagger_{k_2d_2}|0\rangle\\
|\psi_{16}^0\rangle = d_{abc} d_{ad_1e_1}d_{ad_2e_2}d_{ad_3e_3}A^\dagger_{id_1}A^\dagger_{ie_1} A^\dagger_{jd_2}A^\dagger_{je_2}A^\dagger_{kd_3}A^\dagger_{ke_3}|0\rangle\\
\end{array}\label{sp0_states}
\end{eqnarray}
where $f_{abc}$ and $d_{abc}$ are the structure constants of $SU(3)$.  

For spin $1$, there are $10$ different triplets of states, giving a total of $30$ states. Below is the list of all spin-1 states, each state labelled by a free spin index $i$ that takes values from 1 to 3. Since the Hamiltonian is rotationally invariant, all the $3$ states in a given multiplet are degenerate. So it suffices to consider only one state in each multiplet, say, $i=1$.

\begin{eqnarray}
\begin{array}{lll}
|\psi_{1}^1 \rangle= d_{abc}A^\dagger_{jb}A^\dagger_{jc} A^\dagger_{ia}|0\rangle\\
|\psi_2^1\rangle =\epsilon_{jkl}d_{ab_1c_1}f_{ab_2c_2} A^\dagger_{ib_1}A^\dagger_{jc_1} A^\dagger_{kb_2}A^\dagger_{lc_2}|0\rangle\\
|\psi_3^1 \rangle= d_{ace}A^\dagger_{ia}A^\dagger_{jb} A^\dagger_{jb}A^\dagger_{kc} A^\dagger_{ke}|0\rangle\\
|\psi_4^1\rangle = d_{ace}A^\dagger_{ib}A^\dagger_{jb} A^\dagger_{ja}A^\dagger_{kc} A^\dagger_{ke}|0\rangle\\
|\psi_5^1 \rangle= d_{ace}A^\dagger_{ia}A^\dagger_{jb} A^\dagger_{jc}A^\dagger_{ke} A^\dagger_{kb}|0\rangle\\
|\psi_6^1\rangle = d_{abc}f_{bc_1b_2}f_{cc_2b_1} A^\dagger_{ia}A^\dagger_{jb_1} A^\dagger_{jc_1}A^\dagger_{kb_2} A^\dagger_{kc_2}|0\rangle\\
|\psi_7^1 \rangle= \epsilon_{jkl}d_{abc}f_{ade} A^\dagger_{ib}A^\dagger_{jc} A^\dagger_{kd}A^\dagger_{le}A^\dagger_{i_1a_1} A^\dagger_{i_1a_1}|0\rangle\\
|\psi_8^1\rangle = \epsilon_{jkl} d_{ab_1c_1} f_{aa_2b_2} A^\dagger_{ia_1}A^\dagger_{i_1a_1} A^\dagger_{i_1b_1}A^\dagger_{jc_1}A^\dagger_{ka_2} A^\dagger_{lb_2}|0\rangle\\
|\psi_9^1 \rangle= \epsilon_{ijk} d_{ab_1c_1} d_{aa_2b_2} A^\dagger_{ja_1}A^\dagger_{i_1a_1} A^\dagger_{i_1b_1}A^\dagger_{kc_1}A^\dagger_{la_2} A^\dagger_{lb_2}|0\rangle\\
|\psi_{10}^1\rangle = \epsilon_{ijk}d_{ab_1c_1} f_{bb_2c_2}A^\dagger_{i_1b_1}A^\dagger_{i_1c_1} A^\dagger_{la}A^\dagger_{lb}A^\dagger_{jb_2} A^\dagger_{kc_2}|0\rangle\\
\end{array}\label{sp0_states}
\end{eqnarray}

For spin $2$, there are $18$ different multiplets of 5 states each, giving a total of $90$ states. They are arranged as components of a rank-$2$ traceless symmetric tensor, with two indices $i,j$. Again it is sufficient to consider only one component, and we have chosen to work with ${i,j}={1,2}$.

\begin{eqnarray}
\begin{array}{lll}
|\psi_1^2 \rangle= (A^\dagger_{ia}A^\dagger_{ja}-\frac{1}{3}\delta_{ij}A^\dagger_{la}A^\dagger_{la})|0\rangle\\
|\psi_2^2\rangle =  A^\dagger_{i_1a_1}A^\dagger_{i_1a_1} (A^\dagger_{ia_2}A^\dagger_{ja_2}-\frac{1}{3}\delta_{ij} A^\dagger_{i_2a_2}A^\dagger_{j_2a_2})|0\rangle\\
|\psi_3^2 \rangle= (A^\dagger_{ia_1}A^\dagger_{i_1a_1} A^\dagger_{i_1b_1}A^\dagger_{jb_1}-\frac{1}{3}\delta_{ij}A^\dagger_{la_1}A^\dagger_{i_1a_1} A^\dagger_{i_1b_1}A^\dagger_{lb_1})|0\rangle\\
|\psi_4^2\rangle = d_{abc}d_{ade}A^\dagger_{i_1b}A^\dagger_{i_1c} (A^\dagger_{id}A^\dagger_{je}-\frac{1}{3}\delta_{ij}A^\dagger_{ld}A^\dagger_{le} )|0\rangle\\
|\psi_5^2 \rangle= A^\dagger_{i_1a_1}A^\dagger_{i_1a_1} (A^\dagger_{ia}A^\dagger_{ja}-\frac{1}{3}\delta_{ij}A^\dagger_{la}A^\dagger_{la})|0\rangle\\
|\psi_6^2\rangle = \frac{1}{2}d_{abc}(\epsilon_{ikl}A^\dagger_{ja_1}A^\dagger_{ka_1}+\epsilon_{jkl}A^\dagger_{ia_1}A^\dagger_{ka_1})A^\dagger_{la}A^\dagger_{mb}A^\dagger_{mc}|0\rangle\\
|\psi_7^2 \rangle= \frac{1}{2}d_{abc}(\epsilon_{ikl}A^\dagger_{ja}+\epsilon_{jkl}A^\dagger_{ia})A^\dagger_{kb}A^\dagger_{la_1}A^\dagger_{ma_1}A^\dagger_{mc}|0\rangle\\
|\psi_8^2\rangle = \epsilon_{klm}f_{abc} d_{da_1a} d_{da_2b_2}A^\dagger_{ka_1}A^\dagger_{lb} A^\dagger_{mc}(A^\dagger_{ia_2}A^\dagger_{jb_2}-\frac{1}{3}\delta_{ij}A^\dagger_{i_2a_2}A^\dagger_{i_2b_2})|0\rangle\\
|\psi_9^2 \rangle=  A^\dagger_{i_1a_1}A^\dagger_{i_1a_1} A^\dagger_{i_2a_2}A^\dagger_{i_2a_2} (A^\dagger_{ia}A^\dagger_{ja}-\frac{1}{3}\delta_{ij}A^\dagger_{la}A^\dagger_{la})|0\rangle\\
|\psi_{10}^2\rangle = A^\dagger_{i_1a_1}A^\dagger_{i_1a_1} A^\dagger_{i_2a_2}A^\dagger_{i_2a_1} (A^\dagger_{ia}A^\dagger_{ja}-\frac{1}{3}\delta_{ij}A^\dagger_{la}A^\dagger_{la})|0\rangle\\
|\psi_{11}^2\rangle= d_{ab_1c_1}d_{ab_2c_2}A^\dagger_{i_1b_1}A^\dagger_{i_1c_1} A^\dagger_{i_2b_2}A^\dagger_{i_2c_2} (A^\dagger_{ia}A^\dagger_{ja}-\frac{1}{3}\delta_{ij}A^\dagger_{la}A^\dagger_{la})|0\rangle\\
|\psi_{12}^2\rangle = A^\dagger_{i_1a_1}A^\dagger_{i_1a_1} (A^\dagger_{ia_2}A^\dagger_{i_2a_2}A^\dagger_{i_2b_2} A^\dagger_{jb_2}-A^\dagger_{la_2}A^\dagger_{i_2a_2}A^\dagger_{i_2b_2} A^\dagger_{lb_2})|0\rangle\\
|\psi_{13}^2\rangle= d_{aa_2b_2} d_{ac_2e_2} A^\dagger_{i_1a_1}A^\dagger_{i_1a_1} A^\dagger_{i_2a_2}A^\dagger_{i_2b_2}(A^\dagger_{ic_2} A^\dagger_{jd_2}-\frac{1}{3}\delta_{ij}A^\dagger_{lc_2} A^\dagger_{ld_2})|0\rangle\\
|\psi_{14}^2\rangle = \frac{1}{2}(\epsilon_{ikl} A^\dagger_{jb}A^\dagger_{kb}+\epsilon_{jkl} A^\dagger_{ib}A^\dagger_{kb}) \epsilon_{mnp}d_{ab_1c_1}f_{bb_2c_2} A^\dagger_{lb_1}A^\dagger_{mc_1}A^\dagger_{nb_2} A^\dagger_{pc_2}|0\rangle\\

|\psi_{15}^2\rangle= d_{ab_1c_1}d_{ab_2c_2} A^\dagger_{lb_1}A^\dagger_{lc_1} A^\dagger_{mb_2}A^\dagger_{mc_2}(\frac{1}{2} (A^\dagger_{ia} A^\dagger_{jb}+A^\dagger_{ja} A^\dagger_{ib})-\frac{1}{3} \delta_{ij}A^\dagger_{la_2} A^\dagger_{lc_2})|0\rangle\\

|\psi_{16}^2\rangle = d_{ab_1c_1}d_{bb_2c_2}A^\dagger_{i_1a}A^\dagger_{i_1b} A^\dagger_{j_1b_1}A^\dagger_{j_1c_1}(A^\dagger_{ia} A^\dagger_{jb}-\frac{1}{3}\delta_{ij}A^\dagger_{la} A^\dagger_{lb})|0\rangle\\

|\psi_{17}^2\rangle = d_{aa_2b_2}d_{bc_2a_1}A^\dagger_{i_1a_1}A^\dagger_{i_1a_2}A^\dagger_{j_1b_2} A^\dagger_{j_1c_2}  (A^\dagger_{ia}A^\dagger_{jb}-\frac{1}{3}\delta_{ij}A^\dagger_{la}A^\dagger_{lb}) |0\rangle\\

|\psi_{18}^2\rangle= d_{ab_1c_1}d_{aa_2b_2}f_{bb_2c_2} A^\dagger_{i_1b_1}A^\dagger_{i_1c_1} A^\dagger_{i_2c_2}A^\dagger_{i_2d_2}(A^\dagger_{ia_2} A^\dagger_{je_2}-\frac{1}{3}\delta_{ij}A^\dagger_{la_2} A^\dagger_{le_2})|0\rangle\\
\end{array}\label{sp0_states}
\end{eqnarray}

{Note that these states are not orthonormal. However, the discussion of variational calculation can be extended to include a non-orthonormal basis, as outlined in the previous section.}

\subsection{Numerics}

To calculate the matrix elements of $H$ it is useful to express the Hamiltonian in terms of the creation and annihilation operator in the anti-normal ordered form.

In the oscillator basis, the number operator is defined as
\begin{equation}
N=A^\dagger_{ia}A_{ia}
\end{equation}
The Hamiltonian is a sum of $H_0$, the cubic interaction and the quartic interaction,
\begin{equation}
H=H_0+H_c+H_q
\end{equation}

The harmonic oscillator Hamiltonian, $H_0$ can be expressed as
\begin{equation}
H_0=\frac{1}{R}(N+12),
\end{equation}
12 being the total zero-point energy of the 24 oscillators. The cubic interaction Hamiltonian is given by
\begin{equation}
H_c=-\frac{g}{4\sqrt{2}R}\epsilon_{ijk}f_{abc}(A^\dagger_{ia}A^\dagger_{jb}A^\dagger_{kc}+3A_{ia}A^\dagger_{jb}A^\dagger_{kc}+3A_{kc}A_{jb}A^\dagger_{ia}+A_{ia}A_{jb}A_{kc})
\end{equation}
and the quartic term is given by
\begin{align}
H_q &=\frac{g^2}{16R}\left[(A_{ib}A_{jc}A_{id}A_{je}+h.c.)+4(A_{je}A_{id}A_{jc}A^\dagger_{ib}+h.c.)\right.\nonumber\\
&\left.+2(A_{ib}A_{jc}A^\dagger_{id}A^\dagger_{je}+A_{ib}A_{id}A^\dagger_{jc}A^\dagger_{je}+A_{ib}A_{je}A^\dagger_{id}A^\dagger_{jc})\right]\nonumber\\
&-\frac{3g^2}{4R}(A_{ia}A_{ia}+A^\dagger_{ia}A^\dagger_{ia}+2A_{ia}A^\dagger_{ia})+\frac{9g^2}{R}
\end{align}

It is useful to present the Hamiltonian in the anti-normal ordered form because then the computation of matrix element of $H$ between any two variational states will be reduced to evaluating contractions of the form
\begin{equation}
\langle 0| A_{i_1a_1}...A_{i_n a_n} A^\dagger_{j_1b_1}...A^\dagger_{j_n b_n}|0\rangle= \delta_{i_1 j_1}\delta_{a_1b_1}...\delta_{i_n j_n}\delta_{a_nb_n}+\text{permutations of } \{j_r, b_r\}.
\end{equation}
This contraction can be easily defined as a numerical algorithm and used in the computation of all the elements of the matrix $\widetilde{H}$.


\section{References}
\begin{thebibliography}{99}

{\bibitem{Morningstar:1999ff} 
  C.~J.~Morningstar and M.~J.~Peardon,
  Phys.\ Rev.\ D {\bf 56}, 4043 (1997);~Phys.\ Rev.\ D {\bf 56}, 034509 (1999);
%
  Y.~Chen {\it et al.},
  Phys.\ Rev.\ D {\bf 73}, 014516 (2006).
 
 }

\bibitem{Durr:2008zz} 
  S.~Durr {\it et al.},
  Science {\bf 322}, 1224 (2008).

\bibitem{Balachandran:2014iya} 
  A.~P.~Balachandran, S.~Vaidya and A.~R.~de Queiroz,
  Mod.\ Phys.\ Lett.\ A {\bf 30}, no. 16, 1550080 (2015).


\bibitem{Balachandran:2014voa} 
  A.~P.~Balachandran, A.~de Queiroz and S.~Vaidya,
  Int.\ J.\ Mod.\ Phys.\ A {\bf 30}, no. 09, 1550064 (2015)

\bibitem{gribov} 
  V.~N.~Gribov,
  In *Nyiri, J. (ed.): The Gribov theory of quark confinement* 24-38

\bibitem{singer} 
  I.~M.~Singer,
  Commun.\ Math.\ Phys.\  {\bf 60}, 7 (1978).

\bibitem{narasimhan} 
  M.~S.~Narasimhan and T.~R.~Ramadas,
  Commun.\ Math.\ Phys.\  {\bf 67}, 121 (1979).
  
{\bibitem{DeGrand:1975cf} 
  T.~A.~DeGrand, R.~L.~Jaffe, K.~Johnson and J.~E.~Kiskis,
  Phys.\ Rev.\ D {\bf 12}, 2060 (1975).}


\bibitem{Brunner} 
  F.~Brunner, D.~Parganlija and A.~Rebhan,
  Phys.\ Rev.\ D {\bf 91}, no. 10, 106002 (2015).
  
\bibitem{Chen} 
  Y.~Chen and M.~Huang,
  arXiv:1511.07018 [hep-ph].
\bibitem{lattice1}
G.S. Bali, K. Schilling, A. Hulsebos, A.C. Irving, C. Michael and P.W. Stephenson,
Phys. Lett. B  {\bf 309}, 378 (1993).
\bibitem{lattice2}
P. v. Baal and A. S. Kronfeld, 
Nucl. Phys. B {\bf 9},  227 (1989).
\bibitem{lattice3}
C. Michael and M. Teper,
Nucl. Phys. B {\bf 314}  347 (1989).
\bibitem{lattice4}
K. Ishikawa, A. Sato, G. Schierholz and M. Teper
Z. Phys. C {\bf 21},167 (1983).
\bibitem{Savvidy:1982jk} 
  G.~K.~Savvidy,
  Nucl.\ Phys.\ B {\bf 246}, 302 (1984).

\bibitem{Ishikawa1} 
  K.~Ishikawa, G.~Schierholz and M.~Teper,
  Z.\ Phys.\ C {\bf 19}, 327 (1983).

\bibitem{mahul1} 
  M.~Pandey and S.~Vaidya,
  J.\ Math.\ Phys.\  {\bf 58}, no. 2, 022103 (2017).
  
  \bibitem{fermion_apsv}
N.~Acharyya, M. Pandey, S. Sanyal and S. Vaidya, 
under preparation.


\bibitem{Durr:2008rw} 
  S.~Durr {\it et al.},
  Phys.\ Rev.\ D {\bf 79}, 014501 (2009)
  
  \bibitem{Davies:2003ik} 
  C.~T.~H.~Davies {\it et al.},
  Phys.\ Rev.\ Lett.\  {\bf 92}, 022001 (2004)
  
  \bibitem{Aoki:2008sm} 
  S.~Aoki {\it et al.},
  Phys.\ Rev.\ D {\bf 79}, 034503 (2009).
  
  \bibitem{Bernard:2001av} 
  C.~W.~Bernard {\it et al.},
  Phys.\ Rev.\ D {\bf 64}, 054506 (2001).
  
  \bibitem{Aubin:2004wf} 
  C.~Aubin {\it et al.},
  Phys.\ Rev.\ D {\bf 70}, 094505 (2004).
  
  
  \bibitem{Ukita:2007cu} 
  N.~Ukita {\it et al.},
  PoS LAT {\bf 2007}, 138 (2007).
  
\bibitem{Antonio:2006px} 
  D.~J.~Antonio {\it et al.},
  Phys.\ Rev.\ D {\bf 75}, 114501 (2007).
  
  \bibitem{Alexandrou:2008tn} 
  C.~Alexandrou {\it et al.},
  Phys.\ Rev.\ D {\bf 78}, 014509 (2008).
  
\bibitem{Noaki:2007es} 
  J.~Noaki {\it et al.} [JLQCD Collaboration],
  PoS LAT {\bf 2007}, 126 (2007).
  
%

\end{thebibliography}
\end{document}